\documentclass[pdflatex,sn-mathphys-num]{sn-jnl}

\usepackage{graphicx}%
\usepackage{multirow}%
\usepackage{amsmath,amssymb,amsfonts}%
\usepackage{amsthm}%
\usepackage{mathrsfs}%
\usepackage[title]{appendix}%
\usepackage{xcolor}%
\usepackage{textcomp}%
\usepackage{manyfoot}%
\usepackage{booktabs}%
\usepackage{algorithm}%
\usepackage{algorithmicx}%
\usepackage{algpseudocode}%
\usepackage{listings}
\usepackage{float}
\usepackage{geometry}
\usepackage{setspace}
\usepackage{booktabs}
\usepackage{threeparttable}
\usepackage{siunitx}
\usepackage{longtable}
\usepackage{caption}

\theoremstyle{thmstyleone}%
\theoremstyle{thmstyletwo}%
\theoremstyle{thmstylethree}%
\begin{document}

\title[Article Title]{Data-Driven Discovery of Unconventional Antiferromagnets}

\author[1,2,4]{\fnm{Qirui} \sur{Cui}}
\equalcont{These authors contributed equally to this work.}

\author[3]{\fnm{Chenxu} \sur{Liu}}
\equalcont{These authors contributed equally to this work.}

\author*[1,2,5]{\fnm{Anna} \sur{Delin}}\email{annadel@kth.se}

\author*[4,6]{\fnm{Kaiyou} \sur{Wang}}\email{kywang@semi.ac.cn}

\affil[1]{Department of Applied Physics, School of Engineering Sciences, KTH Royal
Institute of Technology, AlbaNova University Center, SE-10691 Stockholm, Sweden}

\affil[2]{Swedish e-Science Research Center, KTH Royal Institute of Technology, SE-10044 Stockholm, Sweden}

\affil[3]{Faculty of Science and Engineering, University of Nottingham Ningbo China, Ningbo 315100, China}

\affil[4]{State Key Laboratory of Superlattices and Microstructures, Institute of Semiconductors, Chinese Academy of Sciences, Beijing 100083, China}

\affil[5]{Wallenberg Initiative Materials Science for Sustainability (WISE), KTH Royal Institute of Technology, SE-10044 Stockholm, Sweden}

\affil[6]{College of Materials Science and Opto-Electronic Technology, University of Chinese Academy of Sciences, Beijing, China}

\abstract{\linespread{1.5} \large 
Unconventional antiferromagnets 
combine zero net magnetization with spin-split electronic bands, offering a distinct, important platform for spintronics. 
Their discovery, however, has so far depended largely on case-by-case studies and on a limited number of compounds with experimentally resolved magnetic structures. 
Here, we overcome these bottlenecks by resolving magnetic ground states across a broad materials database. 
We narrow down 37163 magnets from the Materials Project to 189 collinear antiferromagnets by combining physics-informed prescreening, high-throughput exchange calculations and Luttinger--Tisza analysis. Among these, symmetry analysis identifies 36 altermagnets and 11 Luttinger-compensated ferrimagnets (LCFs), including 22 altermagnets and 9 LCFs that have not been reported previously. 
The identified unconventional antiferromagnets can support nonrelativistic spin Hall effects and doping-tunable spin transport with switchable polarization and giant anisotropy.  
Our framework converts broad structural databases into a curated, symmetry-classified set of experimentally testable compensated spin-split magnets, establishing a scalable route for the efficient discovery of functional antiferromagnets.}
\maketitle

\clearpage
\linespread{1.5} \large 
\section{Introduction}\label{sec1}
Antiferromagnets are promising building blocks for future spintronics applications since they combine several unique advantages, including robustness against external magnetic fields, absence of stray fields, and ultrafast spin dynamics \cite{da1,da2,da3,da4}. 
These features offer a route toward ultrafast, high-density, and low-power-consumption spintronic devices \cite{da5,da6,da7,daa7}. 
Yet, the absence of a net magnetic moment renders antiferromagnets inherently challenging to detect and to harness in practical devices.
Intensive efforts have therefore been devoted to searching or artificially constructing systems that can generate measurable and pronounced spin signals  \cite{da8,da9,da10,da12,daa12,daaa12}. 
However, despite their accuracy, conventional approaches -- whether based on case-by-case experimental characterization or density functional theory (DFT) calculations -- are time-consuming and scale poorly with the rapidly growing number of compounds in modern materials databases.

The emergence of unconventional antiferromagnets that combine vanishing net magnetization with spin-split electronic bands has further highlighted this bottleneck. 
There are two representative 
classes: altermagnets and Luttinger-compensated ferrimagnets (LCFs).
Altermagnets are collinear compensated antiferromagnets whose electronic bands exhibit momentum-dependent spin splitting, because the opposite-spin 
sublattices are connected by crystal symmetries other than pure translation or 
inversion \cite{da13,das13,da14,da15,da16,da17,das18,das19,das20,das21,das22,Bhowal2024,Ahn2019,Noda2016,Okugawa2018,Costa1989,Nathans1963}.
A closely related but physically different class
is that of the LCFs \cite{da18,da19,da20,da21}.
LCFs share the hallmarks of altermagnets, i.e., zero net moment and spin-split bands, but the compensation arises from integer band filling rather than the crystal symmetry connecting opposite-spin sublattices. 
Notably, the identification of both classes requires simultaneous determination of the magnetic ground state and the crystal symmetry relations between opposite-spin sublattices. 
Since both of these quantities are typically unknown for unexplored compounds, and even minor errors in either can lead to misclassifications, large-scale discovery remains particularly challenging.

Existing high-throughput strategies for unconventional antiferromagnets partially address this challenge \cite{da22,da23,Sufyan2026}.  
These strategies take the Bilbao MAGNDATA database \cite{da24} as the candidate pool because it provides experimentally resolved magnetic structures and symmetry information. 
However, MAGNDATA contains only 2287 entries, and the number of compounds is inherently limited by the slow pace of magnetic structure characterization. The result is that many potential promising candidates are overlooked.
By contrast, the Materials Project \cite{da25} provides a far broader candidate space. Its current release comprises DFT calculations for more than 10$^5$ inorganic crystals, based on experimentally reported structures and extended by comprehensive derivative structures. 
However, such a breadth induces a clear limitation for magnet discovery. 
Default high-throughput calculations typically do not explicitly resolve magnetic ground states, because they start from a simple ferromagnetic initialization without systematically exploring different spin configurations \cite{da25,daa25,daa26}.
Notably, even known antiferromagnets can be assigned incorrect magnetic labels, while unconventional candidates may remain entirely hidden. 
Therefore, a scalable route capable of mining unconventional antiferromagnets from the existing, broad structural databases is necessary.

In this work, we establish a high-throughput framework for accurate identification of unconventional antiferromagnets from the broad range of materials present in the Materials Project database. 
By combining physics-informed pre-screening, high-throughput exchange calculations, Luttinger--Tisza (LT) analysis, 
and symmetry classification within one workflow, we can determine the specific magnetic order, as well as whether and how spin splitting emerges in momentum space, without relying on experimentally curated magnetic entries. 
Importantly, from a starting pool of 37163 magnets, the workflow generates a candidate set of 36 altermagnets and 11 LCFs, including 22 and 9 previously unreported members, respectively.
Taking HfFeAs and Co$_2$SiO$_4$ as representative examples, we further show that the identified unconventional antiferromagnets support a nonrelativistic spin Hall effect and various doping-tunable spin transport properties.  
Our framework provides a practical route toward rapid materials discovery for antiferromagnetic spintronics applications. 

\section{Results}\label{sec2} 
\subsection{High-throughput screening workflow}
\begin{figure}[H]
\centerline{\includegraphics[width=1\linewidth]{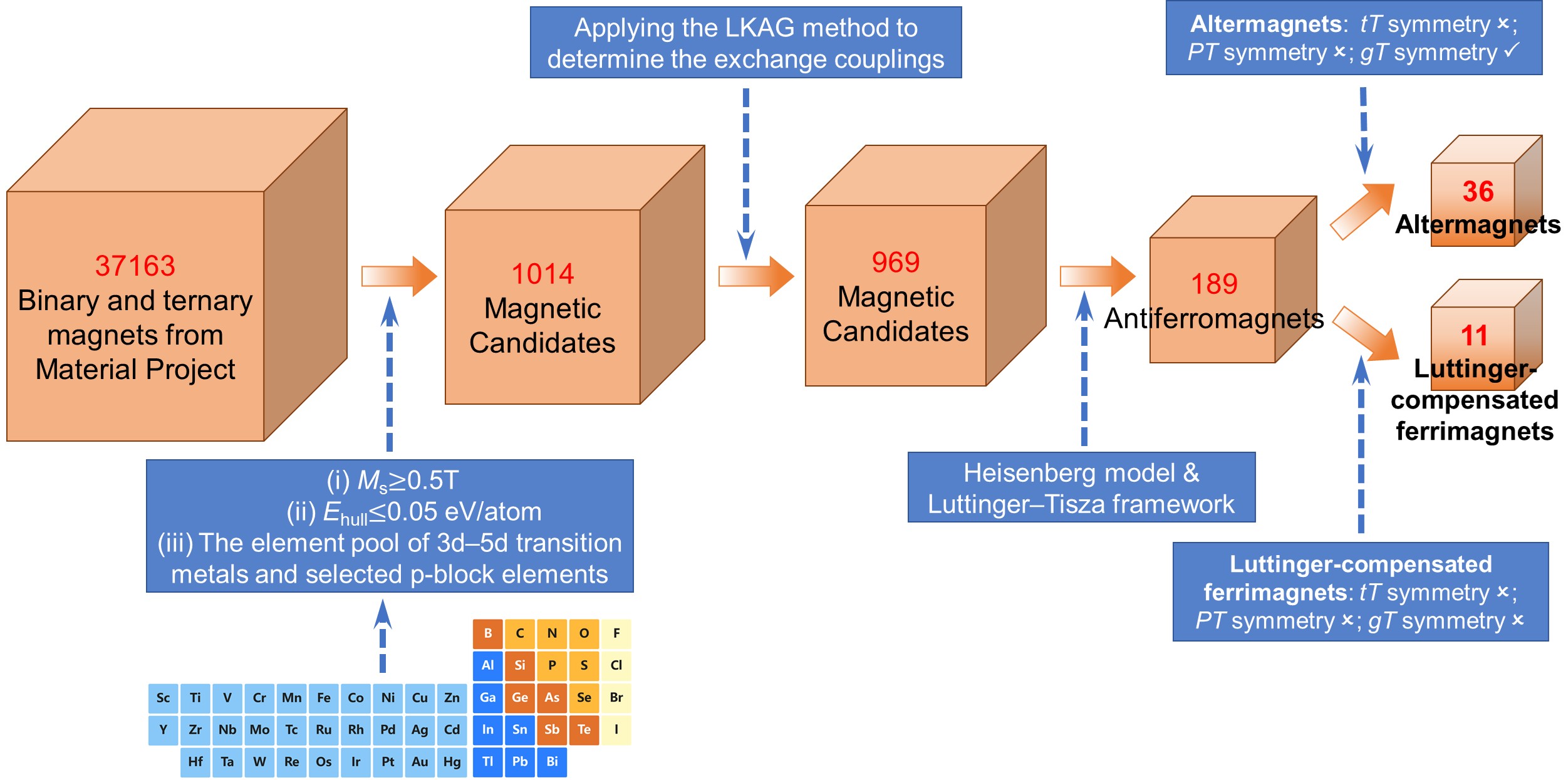}} 
\caption{ 
Schematic high-throughput workflow for the discovery of unconventional antiferromagnets from the Materials Project database. 
The screening starts from 37163 binary and ternary magnetic entries and first applies a physics-informed pre-screening based on three criteria: (i) a finite magnetic moment converged from the default spin-polarized initialization, \(M_s \ge 0.5~\mathrm{T}\); 
(ii) thermodynamic plausibility, enforced by the criterion \(E_{\mathrm{hull}} \le 0.05~\mathrm{eV/atom}\); and 
(iii) the element pool restricted to 3$d$--5$d$ transition metals and selected p-block elements. 
This initial filtering reduces the search space to 1014 magnetic candidates. 
Next, the exchange couplings are extracted for these compounds using the LKAG method, yielding 969 systems with reliable exchange information for subsequent analysis. 
Their magnetic ground states are then determined within the Luttinger--Tisza framework, which identifies 189 compensated antiferromagnets. 
Finally, symmetry analysis is applied to these compensated states to distinguish different classes of unconventional antiferromagnets. 
According to their magnetic symmetry, altermagnets break \(tT\) and \(PT\) symmetries while preserving \(gT\) symmetry, whereas Luttinger-compensated ferrimagnets break all three. 
This procedure yields 36 altermagnets and 11 Luttinger-compensated ferrimagnets.
}
\label{Fig1}
\end{figure}

Figure~\ref{Fig1} shows the high-throughput screening workflow for determining unconventional antiferromagnets. The screening strategy consists of three successive stages including database pre-screening, exchange model construction, and crystal symmetry-based magnets classification. 
We started with 37163 binary and ternary magnets from the Materials Project. 
In order to restrict the search to appropriate and experimentally realizable systems, we first applied filters on magnetization, thermodynamic stability, and composition, retaining only compounds with $E_\mathrm{hull}\le0.05$~eV/atom, $M_s\ge0.5$~T, and elements drawn from 3$d$--5$d$ transition metals together with selected $p$-block elements.
\(E_{\mathrm{hull}} \le 0.05~\mathrm{eV/atom}\) constrains the search within thermodynamically stable or metastable phases~\cite{da27,da28}.
In particular, in the default Materials Project workflow, materials are typically initialized in a ferromagnetic configuration \cite{da25,daa25,daa26}. 
For most entries, the reported magnetic order thus reflects the converged DFT state rather than the ground state. We therefore adopt $M_s \geq 0.5~\mathrm{T}$ as a pragmatic threshold to select compounds that carry appreciable local moments ~\cite{da29,da30}. 
The choice of elemental pool concentrates the search on chemically realistic systems where partially filled $d$ orbitals provide robust local moments, while $p$-block anions and post-transition-metal cations serve as ligands or structural spacers.
This initial step reduces the entire database to 1014 magnetic candidates. 
We then explicitly calculate the spin exchange couplings using the magnetic force theorem \cite{mft1,mft2} and then construct the spin Hamiltonians. 
In total, 969 compounds are carried forward to spin model analysis.  Solving these models within the LT framework \cite{da31,da32} identifies 189 antiferromagnets, while the remaining compounds are classified into 597 ferromagnets and ferrimagnets, and 183 noncollinear magnets. 

Next, we deduce in detail how the spin configurations are determined. 
Based on the DFT-resolved spin exchange, we map each compound onto a spin Hamiltonian, 
\begin{equation}
 \mathcal{H} = -\frac{1}{2}
\sum_{\mathbf{R}}
\sum_{i,j=0}^{N-1}
J_{ij}(\mathbf{R})\,
\mathbf{S}_{ i}(\mathbf{0})\!\cdot\!\mathbf{S}_{j}(\mathbf{R}),    
\end{equation}
where
\(
\mathbf{R}=n_{1}\mathbf{a}_{1}+n_{2}\mathbf{a}_{2}+n_{3}\mathbf{a}_{3}
\) runs over all lattice translations and \(i,j\) label the magnetic sites within the primitive cell. In this convention, positive (negative) \(J_{ij}\) favors ferromagnetic (antiferromagnetic) alignment. 
To find the ground state, we use a reciprocal space method based on the LT approach, while strictly ensuring that the magnetic moment at each site remains constant.  
We therefore write 
\begin{equation}
\mathbf{S}_{j}(\mathbf{R})
=
\Re\!\left[
u_{j}(\mathbf{q})\,
e^{i\mathbf{q}\cdot(\mathbf{R}+\mathbf{r}_{j})}\,
(\mathbf{e}_1-i\mathbf{e}_2)
\right], 
\end{equation}
where \(\mathbf{r}_{j}\) is the basis position of sublattice \(j\), and \(
\mathbf{q}=h\mathbf{b}_{1}+k\mathbf{b}_{2}+l\mathbf{b}_{3}
\) is the ordering vector in reciprocal space. 
\(\mathbf{e}_1\) and \(\mathbf{e}_2\) are orthonormal spin-polarization vectors. 
The complex amplitudes \(u_{j}(\mathbf{q})\) encode the intracell relative phases for the compounds containing more than one magnetic atom in unit cell. Since each spin site possesses a definite moment in realistic candidates, we have $u_j(\mathbf q)=S_j\,v_j(\mathbf q),~|v_j(\mathbf q)|=1$. $
\mathbf v=(v_0,\dots,v_{N-1})^{\mathsf T}$
thus contains only the intracell phase information.  
Substituting the above ansatz into \(\mathcal H\), the system energy reads 
\begin{equation}
E(\mathbf q,\mathbf v)
=
-\frac{1}{2}\,
\mathbf v^\dagger K(\mathbf q)\,\mathbf v,    
\end{equation}
where $[K(\mathbf q)]_{ij}
=
S_i\,[J(\mathbf q)]_{ij}\,S_j$ and $[J(\mathbf q)]_{ij}
=
\sum_{\mathbf R}
J_{ij}(\mathbf R)\,
e^{\,i\mathbf q\cdot(\mathbf R+\mathbf r_j-\mathbf r_i)}.$ 
Physically, \(\mathbf q\) describes how the spin pattern varies from one unit cell to the next, whereas \(\mathbf v\) specifies intracell relative phases of different spin sites. The magnetic structure is finally determined by finding the ordering vector \(\mathbf q^\star\) and the corresponding phase vector \(\mathbf v^\star\) that minimize the $E(\mathbf q, \mathbf v)$.

A final symmetry analysis is performed to distinguish the targeted antiferromagnets. 
Among the candidates, 36 compounds satisfy the defined symmetry conditions of altermagnets, namely broken $tT$ and $PT$ symmetries together with at least one preserved $gT$ symmetry. Here $T$ indicates the time-reversal symmetry, and $t$, $P$, and $g$ indicate the translational,  inversion, and crystallographic symmetry operation other than $t$ and $P$, respectively. Meanwhile, the same workflow identified 11 Luttinger-compensated ferrimagnets (LCFs), for which $tT$, $PT$, and $gT$ are all broken.  

\subsection{Calculation of magnetic ground states}
\begin{figure}[H]
\centerline{\includegraphics[width=1\linewidth]{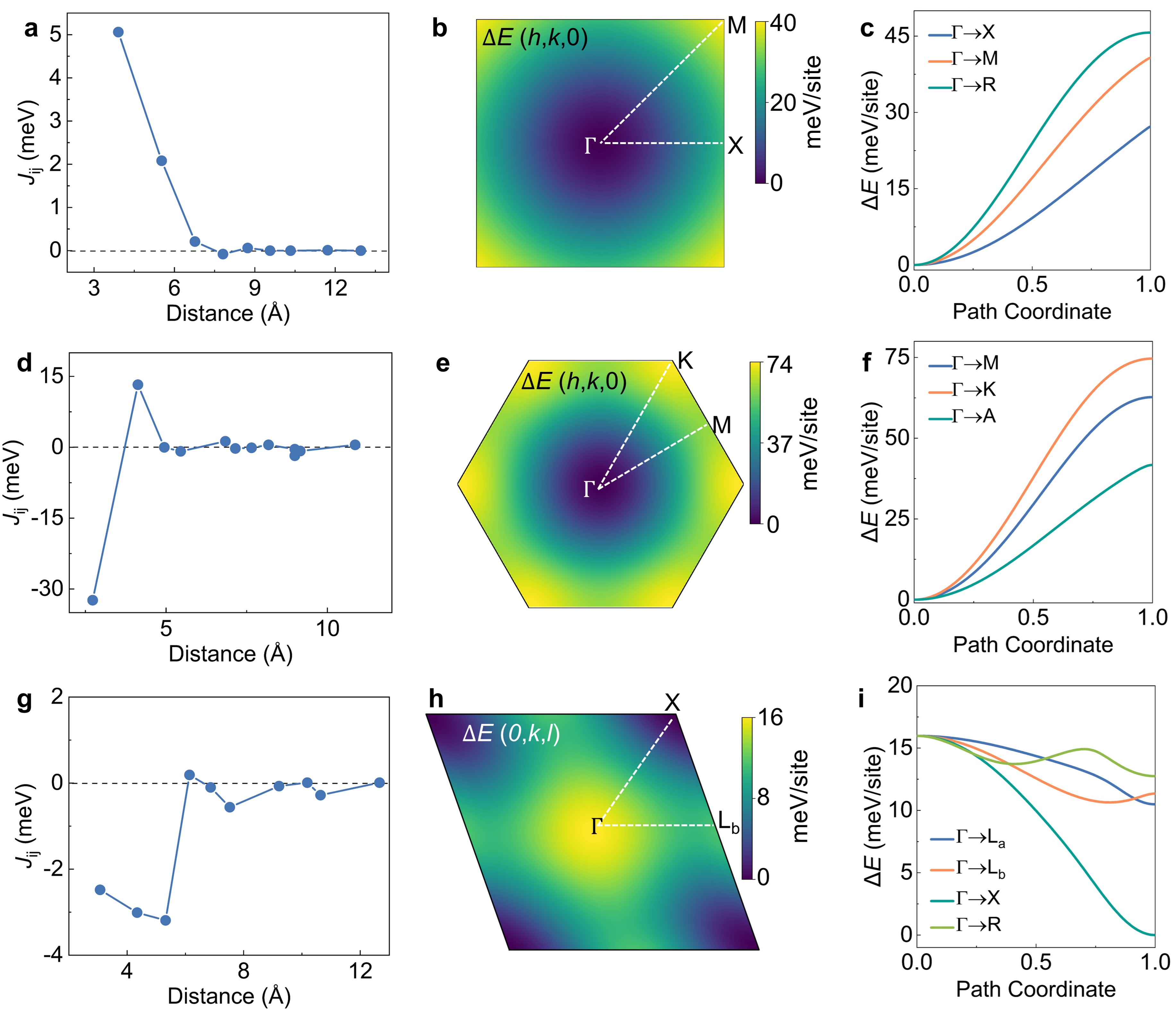}} 
\caption{
Exchange couplings and reciprocal-space magnetic diagnostics for three representative compounds. 
Panels (a--c), (d--f), and (g--i) correspond to CoS$_2$, CrSb, and Mn$_2$PtRh, respectively. 
The left column shows the shell-resolved exchange constants $J_{\text{ij}}$ versus distance. 
The magnitude of exchange couplings approaches zero with increasing spin pair distance in all three candidates, which demonstrates that all dominant exchange couplings are included in our model.
The middle column shows the reciprocal-space energy landscape, $\Delta E(\mathbf q)=E(\mathbf q)-E_{\min}$, in units of meV per magnetic site. 
The right column shows line scans along representative high-symmetry directions. CoS$_2$ and CrSb are minimized at $\Gamma$, consistent with $q=0$ magnetic order, whereas Mn$_2$PtRh is minimized at $\mathbf q^\star_{\rm prim}=(0,1/2,1/2)$. Accordingly, $\Gamma$ is highest in the plotted Mn$_2$PtRh slice, while the $\Gamma\!\to\!$ X scan drops to zero at the endpoint.
}
\label{Fig2}
\end{figure}

Using three representative compounds, CoS$_2$, CrSb and Mn$_2$PtRh, as summarized in Figure ~\ref{Fig2}, we demonstrate how the magnetic ground state is identified in practice based on the exchange model.  CoS$_2$ (space group: \(Pa\bar{3}\)) represents the simplest case. 
As shown in Figure ~\ref{Fig2}\textbf{a}, the exchange is dominated by positive short-range couplings, with the nearest-neighbor Co pairs coupling strongly ferromagnetically with a coupling strength of 5.06 meV in the unit-spin vector convention.  
The subsequent interactions rapidly decaying toward zero. 
This suggests that any deviation from uniform alignment should be penalized. 
Consistent with this expectation, the reciprocal-space energy landscape in Figure ~\ref{Fig2}\textbf{b} has a minimum at the Brillouin zone center, and all line scans in  Figure ~\ref{Fig2}\textbf{c} rise monotonically away from $\Gamma$. The magnetic ordering vector is therefore $\mathbf q^\star=(0,0,0)$, and the crystallographic primitive cell is already the magnetic cell. The corresponding ground state is thus ferromagnetic (Figure S1\textbf{a}). 

The reciprocal-space signature of CrSb (space group: \(P6_3/mmc\)) is deceptively similar to that of CoS$_2$. The spin exchange couplings (Figure ~\ref{Fig2}\textbf{d}) are, however, qualitatively different. Strong antiferromagnetic exchange ($J_{1}=-32.37$ meV) coexists with sizable ferromagnetic
terms ($J_{2}=13.24$ meV) . 
Nevertheless, the reciprocal-space minimum (Figure~\ref{Fig2}e) still emerges at $\Gamma$, and the line scans (Figure~\ref{Fig2}\textbf{f}) confirm that no boundary ordering vector is energetically favored. 
Based on the ordering vector \(\mathbf q^\star\) alone, CoS$_2$ and CrSb would appear identical. 
The distinction instead lies in the sublattice phase of the corresponding eigenvector within the unit cell. 
For a system with \(N\) magnetic sublattices, the optimal eigenvector is defined only up to an overall phase since \(\mathbf v^\star\) and \(\mathbf v^\star e^{i\chi}\) represent the identical spin configuration. 
Therefore, only relative phases are physically meaningful. 
Choosing sublattice $1$ as the reference, we define the mode phase as $
\arg(v_j^\star)-\arg(v_1^\star)$ where $\arg(v^\star)$ denotes the phase angle of  $v^\star$.
The phase determining real-space spin textures is $\phi_j=\mathbf q^\star\!\cdot\!(\mathbf R+\mathbf r_j)+\arg(v_j^\star)$. 
Accordingly, the intracell phase difference is 
\begin{equation}
\Delta\phi_{j1}
=
\mathbf q^\star\cdot(\mathbf r_j-\mathbf r_1)
+
\arg(v_j^\star)-\arg(v_1^\star).
\end{equation}
For CoS$_2$, all four Co spin sites become phase equivalent, so that the \(\Gamma\)-point minimum corresponds to a ferromagnetic state. Interestingly, two Cr spin sites in CrSb exhibit relative phases \((0,\pi)\), i.e., they are exactly out of phase. The resulting state is therefore a collinear antiferromagnet rather than a ferromagnet (Figure S2, row 1, column 2).

Mn$_2$PtRh (space group: \(Fm\bar{3}m\)) illustrates the complementary situation in which both the ordering vector and the phase pattern are decisive to determine the magnetic ground state. 
Figure~\ref{Fig2}\textbf{g} shows that the nearest-neighbor exchange interactions are antiferromagnetic, whereas the longer-range couplings are weaker and partly compensate.  
The energy dispersion
in Figure~\ref{Fig2}\textbf{i} further show that only the \(\Gamma\!\rightarrow\!\)X branch softens continuously toward the global minimum, while the paths toward other high-symmetry points remain at higher energies. 
The magnetic ordering vector is therefore $\mathbf q_{\text{prim}}^\star=(0,1/2,1/2)$. 
Based on Eq. (4), the two Mn moments in primitive cell differ by a phase of \(\pi\) though the optimal eigenvector is identical on two Mn spin sites. Mn$_2$PtRh is thus identified as a collinear antiferromagnet, with a magnetic unit cell doubled relative to the fcc primitive cell (Figure S1\textbf{b}).

The above three materials capture the essential outcomes of the exchange model analysis. In CoS$_2$, the \(\Gamma\)-point minimum and the uniform intracell phase identify a ferromagnet. In CrSb, the same \(\Gamma\)-point minimum instead leads to a collinear antiferromagnet due to its nontrivial intracell phase shift. In Mn$_2$PtRh, the displacement of the minimum away from \(\Gamma\) to X indicates an enlarged magnetic periodicity. The resulting ferromagnetic and antiferromagnetic ground states of CoS$_2$ and CrSb, respectively, are consistent with experimental observation \cite{exp1,exp2,exp3,exp4}.

\subsection{Altermagnets and Luttinger-compensated ferrimagnets}
With the resolved spin configuration, collinear antiferromagnets are classified into four symmetries according to the operations that connect opposite-spin sublattices. The symmetry analysis is performed in the non-relativistic regime. Both symmorphic and non-symmorphic, proper and improper space group operations are considered. 
(i) For candidates with $tT$ symmetry, opposite-spin sublattices are connected by fractional translation $t$. Since $t$ leaves momentum space unchanged, the $T$ symmetry enforces the electronic bands to be spin-degenerate throughout the Brillouin zone.
(ii) Candidates without $tT$ symmetry are then tested for \(PT\) symmetry that enforces spin degeneracy across the Brillouin zone. 
This protection persists even in the presence of spin--orbit coupling, namely Kramers spin degeneracy.
(iii) The remaining candidates are examined for \(gT\) symmetry. 
In the absence of both \(tT\) and \(PT\),  the spin-splitting bands emerge despite zero net magnetization in real space. 
Notably, opposite-spin sublattices can still be connected by rotational, mirror, screw-axis, or glide-plane operations. 
These materials are classified as altermagnets. 
(iv) Candidates failing to meet any of the above criteria possess no SG operation connecting opposite-spin sublattices. 
The opposite-spin sites therefore experience different crystal fields and, in general, carry unequal magnetic moments, which points to ferrimagnetism. However, the net moment may still strictly vanish in insulators and half-metals based on the Luttinger theorem, allowing a compensated ferrimagnetic state. This final class is referred to as the LCF.

Table~I summarizes the 47 candidates identified by our screening workflow, comprising 36 altermagnets and 11 LCFs. The corresponding crystal structures with resolved spin configurations, spin-split electronic bands, and exchange couplings are all provided in the Supporting Information (Figure.~S2--S13). Among the altermagnets, 14 compounds are already known in the literature, including the experimentally confirmed altermagnets MnTe \cite{da16,da17}, CrSb \cite{das18,das19,das20}, $\alpha$-Fe$_2$O$_3$ \cite{das21}, and BiFeO$_3$ \cite{Rbi7,Rbi8,Rbi9}, and the theoretically predicted Fe$_2$PO$_5$ \cite{R8}, YCrO$_3$ \cite{da22,R8}, FeBO$_3$ \cite{da22,R7,R8}, NiCO$_3$ \cite{R7}, CoSO$_4$ \cite{R7}, CoCO$_3$ \cite{R7}, FeF$_3$ \cite{da22,R7,R8}, NiF$_2$ \cite{R8,R9}, TlCrO$_3$ \cite{da22,R7}, and NiTeO$_3$ \cite{R7}. 
On the LCF side, GaFeO$_3$ \cite{R11} and BiNiO$_3$ \cite{R10} have been reported.

Importantly, however, the key advance here is that our workflow does not merely recover some known cases, but substantially expands the accessible material space. Apart from above candidates, we identify 22 previously unreported altermagnets, including Mn$_2$SiS$_4$, MnMoN$_2$, Mn$_2$SiSe$_4$, CoTeO$_3$, Cr$_2$GeC, NiSeO$_3$, TiFeSi$_2$, CoSeO$_3$, Fe$_4$(P$_2$O$_7$)$_3$, Fe$_2$(TeO$_3$)$_3$, Mn$_2$GeS$_4$, Mn$_2$GeSe$_4$, Ti$_2$Fe$_4$O$_9$, CrAuO$_2$, Co(AsO$_2$)$_2$, TiFeP, FeBW, Mn$_2$SnSe$_4$, NbFeSi, HfFeAs, CrFeAs$_{2}$, and Co(SbO$_2$)$_2$, together with 9 previously unreported LCFs, namely AlFeO$_3$, Co$_2$SiO$_4$, Fe(SbO$_2$)$_2$, Nb$_2$Co$_4$O$_9$, Fe$_2$(SeO$_4$)$_3$, Fe$_2$BO$_4$, FeSbO$_4$, MnP$_2$O$_7$, and Cr(PO$_3$)$_2$. 
These results indicate that the screening strategy not only recovers known benchmark systems, but also systematically uncovers new compensated spin-split magnets. 

\begin{table*}[t]
\centering
\captionsetup{font=normalsize,labelfont=bf}
\caption{Unconventional antiferromagnets identified from the high-throughput screening workflow. No. 1---36 are altermagnet candidates, while No.37---47 are LCF candidates.}
\label{tab:materials-summary-47}
\fontsize{9}{10.0}\selectfont
\setlength{\tabcolsep}{3.6pt}
\renewcommand{\arraystretch}{1.2}
\begin{tabular*}{0.9\textwidth}{@{\extracolsep{\fill}} c c c c c c @{}}
\toprule
No. & Materials & MP ID & Space group & Anisotropy & Conduction \\
\midrule
1 & MnTe & mp-404 & P6$_{3}$/mmc & g & I \\
2 & CrSb & mp-1641 & P6$_{3}$/mmc & g & M \\
3 & Mn$_{2}$SiS$_{4}$ & mp-5056 & Pnma & d & I \\
4 & MnMoN$_{2}$ & mp-9374 & P6$_{3}$/mmc & g & I \\
5 & Mn$_{2}$SiSe$_{4}$ & mp-17367 & Pnma & d & I \\
6 & Fe$_{2}$PO$_{5}$ & mp-18307 & Pnma & d & I \\
7 & YCrO$_{3}$ & mp-18725 & Pnma & d & I \\
8 & FeBO$_{3}$ & mp-19097 & R$\bar{3}$c & g & I \\
9 & CoTeO$_{3}$ & mp-19113 & Pnma & d & I \\
10 & NiCO$_{3}$ & mp-19147 & R$\bar{3}$c & g & I \\
11 & CoSO$_{4}$ & mp-19379 & Pnma & d & I \\
12 & Fe$_{2}$O$_{3}$ & mp-19770 & R$\bar{3}$c & g & I \\
13 & Cr$_{2}$GeC & mp-19821 & P6$_{3}$/mmc & g & M \\
14 & NiSeO$_{3}$ & mp-20460 & Pnma & d & I \\
15 & CoCO$_{3}$ & mp-21434 & R$\bar{3}$c & g & I \\
16 & TiFeSi$_{2}$ & mp-21662 & Pbam & d & M \\
17 & FeF$_{3}$ & mp-22398 & R$\bar{3}$c & g & I \\
18 & CoSeO$_{3}$ & mp-22616 & Pnma & d & I \\
19 & BiFeO$_{3}$ & mp-23501 & R3c & g & I \\
20 & NiF$_{2}$ & mp-556324 & Pnnm & d & I \\ 
21 & Fe$_{4}$(P$_{2}$O$_{7}$)$_{3}$ & mp-560126 & P2$_{1}$/c & d & I  \\ 
22 & Fe$_{2}$(TeO$_{3}$)$_{3}$ & mp-605856 & Pnma & d & I \\
23 & Mn$_{2}$GeS$_{4}$ & mp-621925 & Pnma & d & I \\
24 & Mn$_{2}$GeSe$_{4}$ & mp-640047 & Pnma & d & I \\
25 & Ti$_{2}$Fe$_{4}$O$_{9}$ & mp-765583 & Cc & d & I \\
26 & CrAuO$_{2}$ & mp-997159 & P6$_{3}$/mmc & g & I \\
27 & Co(AsO$_{2}$)$_{2}$ & mp-1096874 & P4$_{2}$/mbc & d & I \\
28 & TiFeP & mp-1101870 & Pnma & d & M \\
29 & FeBW & mp-1103093 & Pnma & d & M \\
30 & TlCrO$_{3}$ & mp-1105306 & Pnma & d & I \\
31 & Mn$_{2}$SnSe$_{4}$ & mp-1193446 & Pnma & d & I \\
32 & NiTeO$_{3}$ & mp-1209810 & Pnma & d & I \\
33 & NbFeSi & mp-1211280 & Pmn2$_{1}$ & d & M \\
34 & HfFeAs & mp-1212428 & Pnma & d & M \\
35 & CrFeAs$_{2}$ & mp-1226258 & Pmn2$_{1}$ & d & M \\
36 & Co(SbO$_{2}$)$_{2}$ & mp-2740809 & P4$_{2}$/m & d & I \\
37 & AlFeO$_{3}$ & mp-21216 & Pna2$_{1}$ & s & I \\
38 & Co$_{2}$SiO$_{4}$ & mp-21856 & P2$_1$/c & s & I \\
39 & Fe(SbO$_{2}$)$_{2}$ & mp-22265 & P1 & s & I \\
40 & Nb$_{2}$Co$_{4}$O$_{9}$ & mp-31513 & P1& s & I \\
41 & Fe$_{2}$(SeO$_{4}$)$_{3}$ & mp-505098 & P2$_{1}$/c & s & I \\
42 & Fe$_{2}$BO$_{4}$ & mp-566717 & Pc & s & I \\
43 & FeSbO$_{4}$ & mp-761281 & Imm2 & s & I \\
44 & MnP$_{2}$O$_{7}$ & mp-770531 & P2$_{1}$/c & s & I \\
45 & Cr(PO$_{3}$)$_{2}$ & mp-772172 & C2/c & s & I \\
46 & GaFeO$_{3}$ & mp-868005 & P1 & s & I \\
47 & BiNiO$_{3}$ & mp-1105718 & P$\bar{1}$ & s & I \\
\bottomrule
\end{tabular*}
\end{table*}

\begin{figure}[H]
\centerline{\includegraphics[width=1\linewidth]{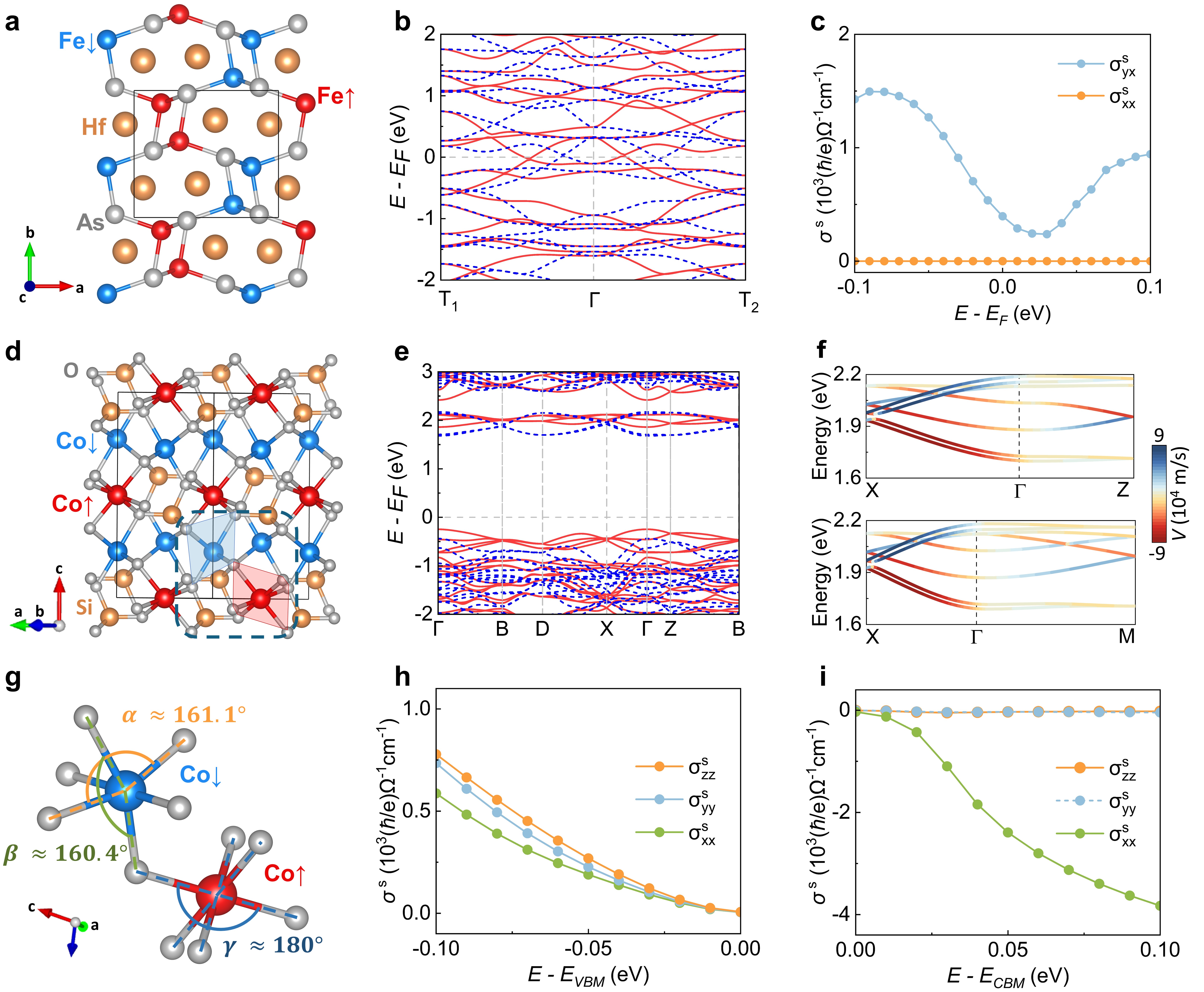}} 
\caption{
Representative identified unconventional antiferromagnets.
(a) Crystal and magnetic structure of HfFeAs. The Fe atoms on the spin-up and spin-down sublattices are shown by red and blue spheres, respectively. 
(b) Spin-split band structure of HfFeAs, where the solid red and dashed blue curves denote the two spin channels. 
(c) The spin conductivity of HfFeAs, showing a finite transverse component and a vanishing longitudinal component. 
(d) Crystal and magnetic structure of Co$_2$SiO$_4$, with the opposite-spin Co sublattices labeled in red and blue. 
(e) Spin-split band structures of Co$_2$SiO$_4$. Different from the \(d\)-wave altermagnetic splitting of HfFeAs, spin splitting emerges throughout the Brillouin zone including the $\Gamma$ point. 
(f) Enlarged view of the conduction bands of Co$_2$SiO$_4$ along X--$\Gamma$--Z and X--$\Gamma$--M, colored by the electronic group velocity. 
The conduction bands exhibit strongly anisotropic dispersion, which leads to pronounced directional dependence of the carrier group velocity. 
(g) Oxygen octahedra surrounding the opposite-spin Co, enlarged from the shadowed region in (d), illustrating the inequivalent chemical environments of opposite-spin sublattices. 
(h,i) longitudinal spin conductivity of Co$_2$SiO$_4$ near the valence band maximum and conduction band minimum, respectively, indicating the doping-tunable spin transport.}
\label{Fig3}
\end{figure}

We next consider the altermagnet HfFeAs and the LCF Co$_2$SiO$_4$ as representative examples to show how the net spin responses emerge in these unconventional compensated antiferromagnets. We find that the altermagnets indeed exhibit the well-known nonrelativistic spin Hall effect, and more interestingly, that the LCFs exhibit various doping-tunable spin transport properties.
The magnetic structure of HfFeAs characterizes a \(d\)-wave altermagnet (Figure \ref{Fig3}\textbf{a}). It consists of Fe dimers with parallel spins, surrounded by nonmagnetic Hf and As atoms, while nearest-neighboring magnetic dimers exhibit opposite spin orientations. 
The \(d\)-wave altermagnetism is characterized by two orthogonal time-reversal glide mirrors normal to the \(a\) and \(b\) axes.
In FeHfAs, the two opposite-spin sublattices are connected by \(C_{2z}T\) symmetry. Due to the anisotropic spin-split band structures (Figure~\ref{Fig3}\textbf{b}), an electric field applied along either \(a\) or \(b\) axis can generate a transverse pure spin current, which is a typical transport property of $d$-wave altermagnets. As shown in (Figure~\ref{Fig3}\textbf{c}), the finite transverse spin current with the magnitude of $395~(\hbar/e)\,\Omega^{-1}\,\mathrm{cm}^{-1}$ emerges, and it can be further enhanced to $1495~(\hbar/e)\,\Omega^{-1}\,\mathrm{cm}^{-1}$ by lowering the Fermi level ($E_{\text{F}}$) by 0.09 eV. By contrast, the longitudinal spin current remains strictly zero.

For Co$_2$SiO$_4$ (Figure~\ref{Fig3}\textbf{d}), the oxygen octahedra surrounding Co sites with opposite spin polarization are manifestly inequivalent. As illustrated in Figure~\ref{Fig3}\textbf{g}, the selected O--Co--O angles of Co$_{\downarrow}$ are both around 160°, indicating the local inversion symmetry breaking. However, the corresponding angles of Co$_{\uparrow}$ remain 180°.  
The distinct local environments of Co$_{\uparrow}$ and Co$_{\downarrow}$ result in the absence of crystal symmetry operations connecting opposite-spin sublattices. This is the core requirement of the LCF. 
Consequently, substantial spin splitting emerges throughout the Brillouin zone (Figure~\ref{Fig3}\textbf{e}). 
Interestingly, the hole and electron doping is able to select not only opposite spin channels but also support fully spin-polarized longitudinal currents. 
The longitudinal spin conductivities near the valence band maximum (VBM) and conduction band minimum (CBM) are shown in Figure~\ref{Fig3}\textbf{h} and \ref{Fig3}\textbf{i}, respectively. 
Specifically, $\sigma_{xx}^{s}=-2393~(\hbar/e)\,\Omega^{-1}\,\mathrm{cm}^{-1}$ when $E_{\text{F}}$ lies at $E_{\mathrm{CBM}}+0.05~\mathrm{eV}$, whereas $\sigma_{xx}^{s}=189~(\hbar/e)\,\Omega^{-1}\,\mathrm{cm}^{-1}$ when $E_{\text{F}}$ is shifted to $E_{\mathrm{VBM}}-0.05~\mathrm{eV}$. 
We also find that the spin conductivity is highly anisotropic near the CBM. 
The transport along $x$ (the crystallographic $a$ axis) is dominant, while responses along $y$ and $z$ directions almost vanished. The anisotropy ratio $(\sigma_{xx}^{s}-\sigma_{zz}^{s})/\sigma_{zz}^{s}$ reaches $4.5\times10^{3}\%$ at $E_{\mathrm{CBM}}+0.05~\mathrm{eV}$. 
This giant anisotropy originates from the strongly reduced conduction-bands dispersion along $\Gamma$---Z, relative to $\Gamma$---X, which suppresses the electron group velocity. 
The maximum $v_z$ is only around $1.1\times10^{4}~\mathrm{m\,s^{-1}}$, almost one order of magnitude smaller than the maximum $v_x$ of $9.7\times10^{4}~\mathrm{m\,s^{-1}}$ (Figure~\ref{Fig3}\textbf{f}). 
FeSbO$_4$ is another interesting LCF that exhibits efficient doping-tunable spin transport. 
The valence bands near the VBM are fully spin-up-polarized, whereas the conduction bands near the CBM remain nearly spin degenerate (Figure S9, row 3, column 1). Consequently, the spin signal is strongly suppressed under electron doping. Specifically, we obtain $\sigma_{yy}^{s}=1342~(\hbar/e)\,\Omega^{-1}\,\mathrm{cm}^{-1}$ when $E_{\text{F}}$ lies at $E_{\mathrm{VBM}}-0.05~\mathrm{eV}$. The magnitude of $\sigma_{yy}^{s}$ then decreases markedly to only $70~(\hbar/e)\,\Omega^{-1}\,\mathrm{cm}^{-1}$ when $E_{\text{F}}$ is shifted to $E_{\mathrm{CBM}}+0.05~\mathrm{eV}$.

The representative examples above allow us to benchmark some promising candidates for spin transport. 
For $d$-wave altermagnets, the nonrelativistic spin Hall 
effect is symmetry-allowed and arises from the momentum-dependent spin splitting. Among the candidates, TiFeSi$_2$, TiFeP, FeBW, NbFeSi, HfFeAs, and CrFeAs$_2$ are metallic $d$-wave altermagnets. 
Because their Fermi levels intersect spin-split bands, these materials can generate electrically driven pure transverse spin currents without additional carrier injection. 
In contrast, the identified LCFs are all insulating. 
Therefore, their spin-polarized electronic transport requires controlled carrier injection, such as electrostatic gating or chemical doping \cite{rass2, ras1,ras2}. 
A material is selected when the absolute spin polarization of density of states (DOS), $|P(E)|$, remains at l00\% over a continuous 0.10 eV window adjacent to 
either the VBM or CBM. The 0.10 eV window is chosen to represent the thermal occupation broadening at room temperature (see $screening~criteria$ in SI).
Eight LCFs are identified by applying above criteria to the spin-resolved DOS in Figure~S15--S17. 
Fe$_2$(SeO$_4$)$_3$ and Nb$_2$Co$_4$O$_9$ satisfy the criteria on the CB side, while FeSbO$_4$, Cr(PO$_3$)$_2$, and GaFeO$_3$ qualify on the VB side. 
AlFeO$_3$ and Co$_2$SiO$_4$ satisfy the criteria on both sides of band gap. In these two compounds, the signs of $P(E)$ at the VBM and CBM are opposite, indicating that hole and electron doping selects opposite spin channels. 

\section{Conclusions and discussions}\label{sec2} 
In summary, we establish a high-throughput route for discovering unconventional antiferromagnets  from a broad structure database rather than from curated magnetic entries. 
In this workflow, we reduce 37163 candidates from Materials Project down to 189 collinear antiferromagnets and finally to 36 altermagnets and 11 LCFs. Apart from recovering benchmark systems such as MnTe, CrSb, $\alpha$-Fe$_2$O$_3$, and BiFeO$_3$, the workflow substantially expands the known landscape by identifying 22 previously unreported altermagnets and 9 previously unreported LCFs.
The identified materials possess testable spin-transport features.
For example, HfFeAs is a $d$-wave altermagnet that can generate transverse pure spin current, and Co$_2$SiO$_4$ and FeSbO$_4$ representing LCFs exhibit doping-tunable longitudinal spin transport, including polarization reversal and giant anisotropy. 
In this sense, the present framework turns large structure databases into an actionable platform for antiferromagnetic spintronics, rather than a passive catalog of candidate compounds.

Our screening uses isotropic Heisenberg exchange parameters obtained in the nonrelativistic limit. Spin--orbit-driven terms such as Dzyaloshinskii--Moriya interactions and single-ion anisotropy are not included at the classification stage. 
This approximation is appropriate for the high-throughput discovery, since the exchange normally dominates the leading magnetic energy hierarchy and the competition between ferromagnetic and antiferromagnetic orders. 
Relativistic interactions typically perform as higher-order corrections, selecting the spin orientation through magnetocrystalline anisotropy or inducing weak spin canting through Dzyaloshinskii–Moriya interactions~\cite{das13,ras3,ras4,ras5,ras6,ras7}.
Moreover, the characteristic spin splitting in altermagnets and LCFs is a nonrelativistic phenomenon, independent of spin-orbit coupling.
The workflow therefore balances the physical fidelity and computational efficiency for the discovery of unconventional antiferromagnets. 

\newpage

\newpage

\newpage

\newpage
\section{Methods}\label{sec4}
\subsection{Exchange coupling calculations}
The Liechtenstein--Katsnelson--Antropov--Gubanov (LKAG) formalism, derived from real-space Green’s functions and the magnetic force theorem, is applied to compute the exchange couplings \cite{mft1}. The LKAG expression for the Heisenberg exchange is written as $J_{ij} 
= -\frac{1}{4\pi}
\,\int_{-\infty}^{E_f}
\operatorname{Im}
\Bigl\{
   \operatorname{Tr}
   \Bigl[
      \mathbf{\delta}_{ii}\,G_{ij}(\varepsilon,\mathbf{R})\,\mathbf{\delta}_{jj}\,G_{ij}(\varepsilon,-\mathbf{R})
   \Bigr]
\Bigr\}
\,d\varepsilon,$
where \(\mathbf{R}\) is the lattice vector connecting spin sites \(i\) and \(j\); \(G_{ij}(\varepsilon,\mathbf{R})\) is the real-space Green's function; and \(\boldsymbol{\delta}_{ii} \equiv H_{ii}^{\uparrow}-H_{ii}^{\downarrow}\) denotes the on-site spin-splitting block of the Hamiltonian. 
The exchange calculations are performed using the QuantumATK (QATK) package \cite{s1}, which represents the electronic states as a linear combination of atomic orbitals (LCAO).  
Because the LCAO method utilizes a finite, localized basis set, it requires far fewer basis functions than plane-wave methods \cite{s2, s3}.  This distinction is advantageous for our high-throughput screening involving over one thousand magnetic compounds.
The spin-polarized DFT calculations are performed within Generalized Gradient Approximation (GGA) using the Perdew--Burke--Ernzerhof (PBE) functional,
together with a cutoff energy of 120~Ha and a Monkhorst--Pack \(k\)-point mesh of size \(k_x \times k_y \times k_z\) chosen to satisfy \(k_\alpha a_\alpha \ge 50\) (with \(a_\alpha\) the corresponding lattice constant). The total-energy convergence threshold is set to \(10^{-6}\)~eV. For candidates with relatively localized transition metal \(3d\) states, such as oxides and chalcogenides, the GGA+$U$ method is applied with a $U_{\text{eff}}$ parameter set to 3 eV. On the other hand, no \(U\) correction is introduced for metallic alloys where the \(U\) may overlocalize the \(d\) states.

\subsection{Estimation of magnetic ground states}
To determine the ordering vector \(\mathbf q\) that minimizes the total energy, we first note that for any fixed \(\mathbf q\), minimizing \(E(\mathbf q)\) is equivalent to maximizing \(\mathbf v^{\dagger}K(\mathbf q)\mathbf v\) under the constraints \(|v_i|=1\). 
Introducing Lagrange multipliers for these constraints and varying the corresponding Lagrangian with respect to \(v_i^{*}\) yields the stationary condition \(v_i=\exp\!\big(i\,\arg[(K(\mathbf q)\mathbf v)_i]\big)\). At maximum, the phase of each \(v_i\) is aligned with that of its local effective field \((K\mathbf v)_i\) (the exchange-weighted sum of the neighboring spins).  
We solve this self-consistency condition by means of the projected power iteration scheme \cite{ppi1}.
The \(\mathbf v^{\dagger}K(\mathbf q)\mathbf v\) increases monotonically until convergence. 
Finally, we obtain \(f_{\max}(\mathbf q)=\max_{|v_i|=1}\mathbf v^\dagger K(\mathbf q)\mathbf v\). The magnetic ground state is resolved by choosing \(\mathbf q^\star=\arg [\max_{\mathbf q} f_{\max}(\mathbf q)\)] together with the corresponding phase vector \(\mathbf v^\star=\arg[\max_{|v_i|=1}\mathbf v^\dagger K(\mathbf q^\star)\mathbf v\)].
Compounds whose optimized states are compensated and have lower energy than that of the ferromagnetic state are counted as antiferromagnetic candidates.

\subsection{Spin transport calculations} 
In the absence of relativistic terms, the dc Kubo--Bastin expression for the linear response of a current operator \(\hat{A}\) to an electric field \(E_j\) reads
\begin{equation*}
\sigma_{ij}[\hat{A}]
= -\frac{e\,\hbar}{\pi V}
\sum_{n,m,\mathbf{k}}
\frac{\Gamma^{2}\,
\mathrm{Re}\!\left[
\langle\psi_{n\mathbf{k}}|\hat{A}|\psi_{m\mathbf{k}}\rangle
\langle\psi_{m\mathbf{k}}|\hat{v}_{j}|\psi_{n\mathbf{k}}\rangle
\right]}
{\big[(E_F-\varepsilon_{n\mathbf{k}})^2+\Gamma^2\big]
 \big[(E_F-\varepsilon_{m\mathbf{k}})^2+\Gamma^2\big]}.   
\end{equation*}
Here \(V\) is the crystal volume, \(|\psi_{n\mathbf{k}}\rangle\) and \(\varepsilon_{n\mathbf{k}}\) are Bloch eigenstates and eigenvalues, \(E_F\) is the Fermi energy, $\Gamma=10$ meV is the broadening, and \(\hat{v}_{j}=(1/\hbar)\,\partial \hat{H}/\partial k_j\) is the velocity operator. The spin conductivity tensors are calculated by setting the operator $\hat{A}=\frac{1}{2}\left\{\hat{S}, \hat{v}_i\right\}$. For quantitative evaluation, we construct a Wannier--interpolated tight--binding Hamiltonian from  the first-principles calculations performed employing the Vienna \textit{ab initio} simulation package (VASP) \cite{s65,s66,s67}, and use it to compute band energies and matrix elements \(\langle\psi_{n\mathbf{k}}|\hat{A}|\psi_{m\mathbf{k}}\rangle\) and \(\langle\psi_{m\mathbf{k}}|\hat{v}_{j}|\psi_{n\mathbf{k}}\rangle\) on  a dense \(\mathbf{k}\)-mesh \cite{s71}. In Wannier projections, the outer disentanglement window is chosen to include all bands relevant to the transport, while an inner frozen window is used to preserve the first-principles band dispersions around the Fermi level. 
The resulting Wannier model is validated by direct comparison between the Wannier-interpolated and DFT band structures along high-symmetry paths.

\newpage
\section{Supporting Information}\label{sec4}
Please see the Supporting Information for (i) the crystal and magnetic structures of CoS$_2$ and Mn$_2$PtRh; (ii) the crystal and magnetic structures, electronic band structures, and exchange couplings of the screening-identified 36 altermagnets and 11 LCFs; (iii) the longitudinal spin conductivities of Co$_2$SiO$_4$; and (iv) the spin-polarized density of states nearby the VBM and CBM of 11 LCFs. 

\section{Acknowledgement}\label{sec5}
Financial support from the Swedish Research Council (Vetenskapsrådet, VR; Grant No. 2024-04986), the Knut and Alice Wallenberg Foundation (KAW; Grant No. 2022.0108), and the National Natural Science Foundation of China (Grant No. 12241405 and 12427805) is acknowledged. 
The Wallenberg Initiative Materials Science for Sustainability (WISE) funded by the Knut and Alice Wallenberg Foundation is also acknowledged.
The computations/data handling were enabled by resources provided by the National Academic Infrastructure for Supercomputing in Sweden (NAISS), partially funded by the Swedish Research Council through grant agreement no. 2022-06725.

\newpage
\linespread{1}

\end{document}